\documentclass[11pt,twocolumn]{IEEEtran}
\usepackage[dvips]{graphics}
\begin{document}

\title{Two Models for the Study \\
of Congested Internet Connections\thanks{This work was supported by 
AFOSR grant \# FA95500410319.}}
\author{Ian Frommer, Eric Harder, Brian Hunt, Ryan Lance, Edward Ott, 
James Yorke\thanks{Ian Frommer, Brian Hunt, Ryan Lance, Edward Ott and 
James Yorke are with the University of Maryland,
 College Park, MD 20742, USA, 301-405-1000.  Eric Harder is with the US
 Department of Defense.  The authors can be contacted by email at
 \texttt{orbit@glue.umd.edu}.}}
\date{Version 1.6: \today}
\maketitle

\begin{abstract}
In this paper, we introduce two deterministic models aimed at capturing 
the dynamics of congested Internet connections.  The
 first model is a continuous-time model that combines a system of
 differential equations with a sudden change in one of 
the state variables.  The second model is a discrete-time model with a
time step that arises naturally from the system.  Results from these
models show good agreement with the well-known \emph{ns} 
network simulator, better than the results of a previous, similar  model. 
 This is due in large part to the use of the sudden change to reflect the 
impact of lost data packets.  We also discuss the potential use of this model in 
network traffic state estimation.
%  We also show the results of our models on a network that 
%is prone to irregular behavior due to the presence of a source that sends
% large bursts of packets at regular intervals.  The success of our models 
%in these cases suggests their utilization in a larger scheme to predict 
%network traffic over short time spans.  
\end{abstract}

\section{INTRODUCTION}\label{intro}
An Internet connection consists of the exchange of data packets between 
a source and destination through intermediate computers, known as routers.
  The transmission of data in the majority of Internet connections is 
controlled by the Transport Control Protocol (TCP)~\cite{MAF04,FKM03}.
  TCP is responsible
 for initializing and completing connections, controlling the rate of flow 
of data packets during a connection, ensuring that lost packets are retransmitted,
 etc.  

Congestion can occur when the rate of flow of the connection is limited by
some link on the path from source to destination.  This 
link is known as the \emph{bottleneck} link.  Routers have buffers in which to
 store packets in case one of their outgoing links reaches full capacity.
  If new packets arrive at a router whose buffer is full, that router
 must drop packets.  There exist various strategies, known as active 
queue management (AQM) policies, for determining how and when to drop 
packets prior to the buffer filling.  One commonly used AQM is 
Random Early Detection (RED)~\cite{FJ93}.

In this paper we model the interaction between TCP and RED for a simple
 network connection experiencing congestion.  This scenario, or ones similar 
to it, have been modeled previously for the 
purposes of evaluating RED~\cite{MGT00,FB00,RA01,LPW02}, obtaining 
throughput expressions~\cite{MSMO97,PFTK98}, and obviating the need for 
time-consuming simulation~\cite{MGT00,HBOL01,BHL03,LLM03}, among others.
 There are stochastic~\cite{PFTK98} and deterministic models~\cite{MGT00,FB00,RA01}, 
continuous-time~\cite{MGT00} and 
discrete-time models~\cite{FB00,RA01}. 
% Our intention is to obtain simple deterministic models 

Our models are closest to that of Misra et al.~\cite{MGT00}.  That model
 successfully captures the \emph{average} network state in a 
variety of situations.  This allows the authors to analyze RED from a control
theoretic basis.  Our aim is to develop a model of greater accuracy that will 
be useful in estimation not only of the average network state, 
but of additional
quantities.  For example, a model capable of capturing the mean queue 
length allows one to estimate the mean round-trip time.  But a model that
can capture the range, and better still, the variance of the queue length, will 
allow one to estimate the range and variance of the round-trip time.  
While this level of model accuracy can be useful, it is necessary in
 a model 
that is to be used for our ultimate goal of network traffic state
estimation.  Given some knowledge of the state of the network, an accurate 
model may be combined with a filter-based state-estimation scheme 
(which we discuss briefly in Sec.~\ref{Ongoing}), in order to estimate 
the full network state at the current time.  Network state estimation can
be useful from the perspectives of both security and performance. 
 
We make several assumptions about the Internet connection we are modeling.
  First, we assume it involves the transfer of a
 very large amount of data over an extended period of time.  In this so-called
 \emph{bulk transfer}, the senders always have data to transmit for the
 duration of the connection.  Common TCP traffic such as FTP (File Transfer
Protocol) file transfers,
 and non-streaming music and video downloads can all constitute bulk transfer.
  Second, we assume that the path from source to destination is fixed for the
 duration of the connection.  Third, we assume the transfer is one-way, i.e.,
 the data packets travel in only one direction.  (Acknowledgments packets 
only travel in the other direction.)  Fourth, any cross-traffic through the 
path is negligible.  Fifth, the path contains one bottleneck link whose 
capacity is less than that of all other links in the path.  Most of these assumptions 
reflect typical network scenarios with the possible exceptions of the cross-traffic
and bottleneck assumptions.  

\begin{figure}
 \begin{center}
 \scalebox{.75}{\includegraphics{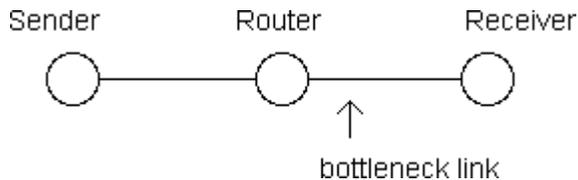}}
 \end{center}
 \caption{Network configuration}
 \label{nc_fig}
\end{figure}

Data traveling from the sender to the receiver is likely to pass through 
several intervening routers.  However, based on our assumption that there is one 
bottleneck link, we need only consider the router whose outgoing link is 
this bottleneck link.  All other routers simply forward the data along the path,
and have unoccupied buffers.  Combining all of the above assumptions, we can model 
the network we are attempting to represent with a network of one sender 
and one receiver separated by one router (see Fig.~\ref{nc_fig}).  

\section{THE MODELS}
The primary mechanism used by TCP to regulate the flow of data is the 
\emph{congestion window} on the sender's computer.  
A window size of $W$ means that the number of
unacknowledged packets from the sender that may be
outstanding on the network at any one time is at most $W$.
  According to TCP, the sender's window size follows the 
additive-increase/multiplicative-decrease (AIMD) scheme.  Typically, 
the window size $W$ increases additively by $1/W$ with each successful
packet transmission and decreases multiplicatively by a factor of $2$ 
when packets are dropped. 

Packets needing to be stored in a router's buffer due to congestion are 
placed in a queue.  The RED module operating at this router keeps track 
of the queue length as well as of an exponentially-weighted average of the
 queue length, which we denote by $x$.   When $x$ exceeds a pre-defined minimum
 threshold, RED will cause the router to drop arriving packets with 
probability that increases with $x$.  Note that this adds a stochastic element
 to the connection.  We attempt to capture the system behavior with a 
deterministic model; our model will not reflect the random component of the 
dynamics.  However, as we will show below, the random component is reasonably 
small, and so a purely deterministic model can be useful for state estimation.

Previous attempts to model this scenario deterministically have made use
of either discrete-time maps~\cite{FB00,RA01} or differential 
equations~\cite{MGT00}.  Continuous-time models employing delay differential
equations may do well in capturing the aspects of network behavior that
evolve on small-time scales, such as flow rate increases.  However, they are
likely to smooth out the effects of sudden large changes in state (such as flow
rate reductions).  Furthermore, systems of
delay differential equations can be cumbersome to express and solve due to
the delay.

We address the above issues through innovations in our models, namely, 
a discrete impulse in the continuous-time
model, the choice of time-step in the discrete-time model, and a
packet-based frame of reference used by both models.  
The continuous-time model presented in this work
combines a system of differential equations to capture the evolution of
the small-time scale changes with a discrete impulse (sudden change in one
of the state variables) to represent sudden large changes. 
The discrete-time model takes advantage of the fact
that a congested bottleneck will impose uniform packet spacing.  Using
this spacing as the time step allows the discrete-time model to nicely
capture the system behavior.  In addition, as explained below, 
both models utilize a
packet-based frame of reference for the state, which simplifies handling of
the delay in the system, making the models easier to express and execute.

To describe the system, we consider the state variables, all of which have 
units of ``packets'' but are allowed to take non-integer values:
\begin{itemize}
\item \textbf{$W$} -- the congestion window size of the sender;
\item \textbf{$q$} -- the length of the queue at the router;
\item \textbf{$x$} -- the exponentially-weighted average of the queue 
length at the router.
\end{itemize}
In our packet-based frame of reference, $W(t)$ represents the congestion 
window in effect for a packet \emph{arriving at the queue} at time $t$. 
This was the sender's congestion window at an earlier time, when the packet 
left the sender.  

In developing the models we will often refer to the round-trip time, $R$.  
The round-trip time is the time between the departure of a packet from the 
sender and the return to the sender of the acknowledgment for that packet.  
It consists of a fixed propagation delay, $a$, and a queuing delay.  For a 
packet encountering a single congested router with queue length $q$, and outgoing link capacity 
$c$ (measured in packets per unit time), the queuing delay is $q/c$.  Thus, 
\begin{equation}\label{R_equa}
R(q)=a+\frac{q}{c}.
\end{equation}

\subsection{THE CONTINUOUS-TIME MODEL}
Our continuous-time model consists of two distinct parts. The first part 
is a system of differential equations similar to the one found 
in~\cite{MGT00}.  It is used to model the additive-increase of the window 
size, and both instantaneous and averaged queue lengths. The second part 
of the model is an impulse in which the window size state variables are 
instantaneously reset.  This is used to model the multiplicative decrease 
in the window size caused by a dropped packet.

\subsubsection{MODELING THE ADDITIVE INCREASE}\label{contAddInc}
According to TCP, the window size, $W$, will normally increase by $1/W$ upon 
receipt of each acknowledgment.  (We do not model the ``slow start''
% \footnote{Apparently ``slow start'' is so named because beginning in slow 
% start, the connection has a small window size, thus sending at a slower 
% rate than during the later ``congestion avoidance'' phase.} 
phase
 in which the window size increases by one per received acknowledgment.) 
 Since $W$ specifies the number of unacknowledged packets that can be out 
in the network at any one time, it 
follows that about W packets are sent per round-trip time, $R$.  Thus, in 
approximately one round-trip time, $W$ acknowledgments return (assuming no 
lost packets), causing $W$ to increase by $1$.  
Although network data is transmitted in discrete packets,
during the period of additive-increase and in the absence of dropped 
packets, quantities appear to vary smoothly when viewed over the time-scales 
we are interested in (on the order of $1$ to $100$ seconds) for our 
network settings. Hence we model the additive-increase of the window size 
continuously:
\begin{equation}\label{W_equa}
\frac{dW(t)}{dt}=\frac{1}{R(q(t))}.
\end{equation}

The rate of change of the queue length at a given time is equal to the 
difference between the 
flows into and out of the router.  By the explanation above, the flow into
 the router at time $t$ is $W(t)/R(q(t))$.  If the queue is occupied, the flow 
rate out will be equal to the capacity of the router's outgoing link, $c$.  
Otherwise the flow out is equal to the minimum of $c$ and the incoming flow 
rate.
\begin{equation}\label{q_equa}
\frac{dq(t)}{dt}=
  \left\{
    \begin{array}{ll}
      \frac{W(t)}{R(q(t))}-c & \mbox{when $q(t)>0$}, \\
      max\left(\frac{W(t)}{R(q(t))}-c,0\right) & \mbox{when $q(t)=0$}.
    \end{array}
  \right.
\end{equation}

The exponentially-weighted average queue is determined by RED upon each 
packet arrival as follows:
\begin{equation}\label{x1_equa}
x_{n+1}=wq_{n+1}+(1-w)x_n,
\end{equation}
where $w$ is the exponential-weighting parameter, $0 \le w \le 1$, and the 
subscripts denote successive measurements of $q$ and $x$ (which occur at 
packet arrivals) for a given router.  This equation describes a low-pass 
filter, and, making use of the fact that $w$ is small in practice, it can be 
approximated by the differential equation:
\begin{equation}\label{x2_equa}
dx(t)/dt = w (q(t)-x(t)) (W(t)/R(q(t))).  
\end{equation}
The term $W(t)/R(q(t))$ expresses the rate of packets arriving at the queue.
The model is summarized by equations ~(\ref{W_equa}),~(\ref{q_equa}), 
and~(\ref{x2_equa}).

Because of our bulk-transfer and bottleneck assumptions 
(see Sec.~\ref{intro}), the network is often congested to the point of 
saturation.  This allows us to simplify the model by discarding the $q=0$ case 
in equation~(\ref{q_equa}), and replacing the packet rate term in 
equation~(\ref{x2_equa}) with the bottleneck link capacity, $c$, as follows:

% \begin{eqnarray}\label{satModel}
\begin{equation}\label{W2_equa}
\frac{dW(t)}{dt}=\frac{1}{R(q(t))}
\end{equation}
\begin{equation}\label{q2_equa}
\frac{dq(t)}{dt}=  \frac{W(t)}{R(q(t))}-c 
\end{equation}
\begin{equation}\label{x3_equa}
\frac{dx(t)}{dt}=wc(q(t)-x(t)).
\end{equation}

\subsubsection{MODELING THE MULTIPLICATIVE DECREASE}
In a real network situation, a router using RED will drop an arriving 
packet with probability $p$ given by:

\begin{equation}\label{RED_equa}
p(x)=
  \left\{
    \begin{array}{lll}
      0 & \mbox{when  $x<q_{min}$}, \\
      \frac{x-q_{min}}{q_{max}-q_{min}}p_{max} & \mbox{when $q_{min} 
          \le x \le q_{max}$ }, \\
      1 & \mbox{when $x > q_{max}$}.
    \end{array}
  \right.
\end{equation}

In practice, this usually causes a packet to be dropped soon after $x$ 
exceeds $q_{min}$.  For the purposes of keeping our model simple and 
deterministic, we assume a drop occurs as soon as $x$ exceeds $q_{min}$.  
We continue to evolve the continuous system for one round-trip time to 
reflect the delay in notification of the sender that a packet has been 
dropped.  Then we cut the window state variable, $W$, in half.   Next 
the continuous evolution resumes but we hold the window state variable 
constant for one round-trip time\footnote{For simplicity, we are neglecting
 the period in which no data are transmitted which would necessarily occur 
after the window cut.}.  This is done in order to represent 
the Fast Recovery/Fast Retransmit(FRFR) algorithm in the 
NewReno\footnote{Recent indications are that NewReno appears to be the most 
widely used variant 
of TCP~\cite{MAF04}.}  version of TCP.  Once this round-trip time has 
elapsed, the model resumes continuous evolution as described in Section 
~\ref{contAddInc}.  

\begin{figure}
 \begin{center}
 \scalebox{.5}{\includegraphics{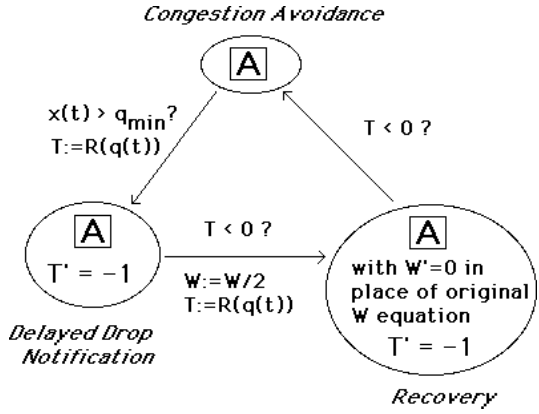}}
 \end{center}
 \caption{Hybrid systems view of the continuous-time model.  \textbf{A} 
stands for equations~(\ref{W2_equa}),~(\ref{q2_equa}), and~(\ref{x3_equa}).
$T'=dT/dt$ and $W'=dW/dt$.}
 \label{hs_fig}
\end{figure}

Denoting the continuous-time system of equations 
~(\ref{W2_equa}),~(\ref{q2_equa}), and~(\ref{x3_equa}) by 
\textbf{A}, we can summarize the continuous-time model in hybrid systems form
as in Fig.~\ref{hs_fig}, where $T$ is a timer variable.  The use of a hybrid
system to express the model was suggested by~\cite{HBOL01}.  Our model begins
in the congestion avoidance phase, following the set of equations, \textbf{A}.
Once $x$ exceeds $q_{min}$, the model assumes a drop occurs, and it 
transitions to the delayed drop notification phase.  In this phase, it 
continues following the original set of equations
but now it also utilizes a count-down timer, $T$, which expires one 
round-trip time from the time $x$ exceeds $q_{min}$.  Once this 
timer expires, the model transitions into the recovery phase, first 
cutting $W$ in half and initializing a new count-down
timer at the value of the present round-trip time.  In the recovery 
phase, the equations still hold, but with one exception:  $W$ is now held 
fixed.  
Finally, when this count-down timer
expires, the model returns to the congestion avoidance phase.

\subsubsection{MODELING MULTIPLE SENDERS}
The continuous-time model can be easily generalized to handle more than
one sender.  As in the case of one sender, we assume each additional sender is 
upstream from the bottleneck link and engaged in bulk transfer, sending
data as fast as allowed by TCP.  Following~\cite{MGT00}, denote the
window size of the $i^{th}$ sender by $W_i$.  We keep track of each sender's
window separately.  The queue equation~(\ref{q_equa}) is replaced by:
\begin{equation}
\frac{dq(t)}{dt}=
  \left\{
    \begin{array}{ll}
      (\sum_{i=1}^N \frac{W_i(t)}{R_i(q(t))})-c & \mbox{when $q(t)>0$}, \\
      max( (\sum_{i=1}^N \frac{W_i(t)}{R_i(q(t))})-c,0) & \mbox{when $q(t)=0$}.
    \end{array}
  \right.
\end{equation}
The connections need not all share the same fixed propagation delay, 
as the $R_i$ terms indicate.  We use a separate copy of~(\ref{W_equa}) for
each window $W_i$, and a separate timer variable for each connection.

\subsection{TOWARDS A DISCRETE-TIME MODEL}
In practice, the differential equations in the continuous-time model can
be solved by numerical integration using a simple scheme such as Euler's
method.  The main
requirement is that the time step be small enough to reflect system
behavior occurring on the smallest time-scale.  In the given network
scenario, the smallest time scale is the spacing between packets, which, as
dictated by the bottleneck link, is equal to $1/c$.  We will now show that
the use of $1/c$ as a time step allows us to express the model in a
simplified way as a discrete-time system.

The Euler numerical integration step for the $W$ equation~(\ref{W_equa}),
is
\begin{equation}\label{WE_equa}
W_{k+1} = W_k + h \frac{1}{a+q_{k}/c},
\end{equation}
where $h$ represents the time step.  Setting $h=1/c$, we have
\begin{equation}\label {WE2_equa}
W_{k+1} = W_k +  \frac{1}{ac+q_k}.
\end{equation}
For the $q$ equation~(\ref{q_equa}), neglecting the $q(t)=0$ term for the
moment, the Euler step is
\begin{equation}\label {qE_equa}
q_{k+1} = q_k + h \left(\frac{W_k}{a+q_{k}/c} - c\right),
\end{equation}
which simplifies to
\begin{equation}\label {qE2_equa}
q_{k+1} = q_k +  \frac{W_k}{ac+q_{k}} - 1,
\end{equation}
using the time step $1/c$.  Note that in the above equation, the middle
term represents the flow in per time step and the final term, the $-1$,
represents the flow out per time step.  In case the queue is empty, we
omit the $-1$ term as no packets will leave within that time step:
\begin{equation}\label{qE3_equa}
q_{k+1}=
\left\{
    \begin{array}{ll}
	q_k +  \frac{W_k}{ac+q_{k}} - 1  & \mbox{when $q_k>0$}, \\
	q_k +  \frac{W_k}{ac+q_{k}} = \frac{W_k}{ac} & \mbox{when $q_k=0$}.
     \end{array}
\right.
\end{equation}
% This equation is more accurate than equation~(\ref{q_equa}).

Finally, it is easy to show that  this choice of time step returns the
equation for $x$ back to its original form.
\begin{equation}\label{xE_equa}
x_{k+1}=wq_{k+1}+(1-w)x_k.
\end{equation}
Note, we are still using the assumption that RED is updating the average
queue size as if packets were arriving at the queue at a rate of $c$.

Summarizing these equations, we have a discrete-time model representation
of the original continuous-time system that gives an intuitive description
of the TCP dynamics on the shortest time-scale.
We can utilize the same overall procedure as the
continuous-time model (see Fig.~\ref{hs_fig}), substituting the 
maps~(\ref{WE2_equa}),~(\ref{qE3_equa}) and~(\ref{xE_equa})
above for the differential equations~(\ref{W2_equa}),~(\ref{q2_equa})
 and~(\ref{x3_equa}).

\subsection{THE DISCRETE-TIME MODEL}
We can derive a similar but even more simplified discrete-time model by
making the assumption that the system is saturated.
As with the continuous-time model, the discrete-time model uses the three
 state variables $W$, $q$, and $x$ and consists of two primary components.
  The first component
 is  a discrete-time map used to model the change in each of the state 
variables during the additive increase phase.  The second part is an impulse 
that works similar to the one in the continuous-time model -- it accounts 
for the sudden adjustment in the window size due to a dropped packet. 

\subsubsection{MODELING THE ADDITIVE INCREASE}
In practice, after an initial transient, the bottleneck link in the scenario 
we model will reach saturation and a queue will form at the router.  Packets
 leave the queue at evenly spaced intervals of length $\delta = 1 /c$, where 
$c$ is the capacity of the outgoing link.  For instance, if the link has a 
capacity of $200$ packets per second, the outgoing packets will be spaced 
(front-to-front) by $1/200 s$ or $.005 s.$\footnote{Throughout this paper we
 assume uniform packet size unless specified otherwise.}  When these
 packets reach their 
destination, acknowledgment packets are sent out, with the same spacing.  
The arrival of the acknowledgment packets at the sender results in the sending
 of new data packets.   Receipt of an acknowledgment frees one packet position 
in the TCP send window, $W$, and causes $W$ to increase by $1/W$.  Typically
this means that one packet is sent for each arriving acknowledgment.  
Occasionally, when $W$ reaches the next integer value, two packets are sent 
upon receipt of an acknowledgment.  
In other words, usually one packet arrives at the
queue every $\delta$ seconds while one packet leaves the queue every $\delta$
seconds, resulting in no net change in queue length.  But occasionally, $2$ 
packets arrive at the queue in a $\delta$-second interval while still only one 
leaves.  This is the primary cause of queue length increase in this scenario.

The description above refers to what is sometimes known as the 
\emph{self-clocking} 
nature of TCP.  That is, a TCP sender has the ability to determine the 
limiting capacity in its flow path and send its packets at that rate.  
We take advantage of this behavior in our model, by employing $\delta$ as 
the time step in a discrete-time model of this system.  

The time step is given by:
\begin{equation}\label{tk_equa}
t_{k}=t_{0}+k\delta, \; k = 0,1,2,\ldots
\end{equation}
where the time $t_0$ marks the start of saturation.  Our equation for $W$ 
follows directly from the rules of TCP:

\begin{equation}\label{Wk_equa}
W_{k+1}=W_{k}+\frac{1}{W_k}.
\end{equation}

We express the change in the queue length as follows:

\begin{equation}\label{qk_equa}
q_{k+1}=q_k+\frac{1}{W_k}.
\end{equation}

While this yields fractional-valued queues, it does result in the queue 
length increasing by $1$ each time $W$ increases
 by $1$, as described above, and 
is simpler to express than the more realistic alternative.  The equation 
for $x$ is the actual equation used by RED (the assumption of saturation 
means packets arrive each time step and so $x$ is updated each time step):

\begin{equation}\label{xk_equa}
x_{k+1}=wq_k+(1-w)x_k.
\end{equation}

The round-trip time, from~(\ref{R_equa}), is:
\begin{equation}\label{Rk_equa}
R(q_k) =  a + \frac{q_k}{c} = \left(\frac{a}{\delta} + q_k\right) \delta.
\end{equation}

In multiples of the time step $\delta$, the round-trip time is thus best 
approximated by $m_k = nint(a/\delta + q_k)$ where ``nint'' means the nearest 
integer.  

We initialize the additive-increase phase of the model at the start of 
saturation, or, in other words, at the onset of queuing.  Just prior to this 
time, there is an exact balance between the flow in and out of the router.  
The flow in is equal to $W(t_0)/R(q(t_0)) = W(t_0)/(a+ q(t_0)/c)$.  Since $q(t_0) 
= 0$ at this point, the flow in is $W(t_0)/a$  and the flow out is $c$. 
Thus the initial conditions for the discrete-time model are $q_0= 0$ 
and $W_0 = ca$.  (Note that $W_0$ corresponds to the so-called 
\emph{bandwidth-delay product} which is often used to estimate buffering
for network resources.)

\subsubsection{MODELING THE MULTIPLICATIVE DECREASE}
We model the multiplicative decrease here in much the same way as in the 
continuous-time model but with one additional element.  As in the 
continuous-time model, we assume a drop occurs as soon as $x$ exceeds 
$q_{min}$:  If $x_k > q_{min}$ then at time $k+m_k-1$, replace~(\ref{Wk_equa})
with
\begin{equation}\label{Wk2_equa}
W_{k+m_k} = \frac{1}{2}W_{k+m_k-1},		
\end{equation}
which reflects the delay of $1$ round-trip time in drop notification.  
We take advantage of the packet-level timing of this model to incorporate
an additional aspect of what occurs to a network sender following a window 
cut.  In a network, after cutting its window from $W$ to $W/2$, the sender will
not send packets for $W/2$ time steps.  This reflects the fact that 
acknowledgments for packets in the half of the window that is no longer
accessible will not trigger the transmission of any new packets.  As a result,
the queue length will decrease.  We model this by using the following 
equations for the next $W/2$ timesteps where $W$ is the window size just prior 
to the window cut:
\begin{equation}\label{Wk3_equa}
W_{k+1} = W_k,
\end{equation}
\begin{equation}\label{qk3_equa}
q_{k+1} = q_k - 1.
\end{equation}
Note that $W_k$ remains fixed following the rules of Fast Recovery/Fast 
Retransmit, as was done in the continuous-time model.  Next we continue
to hold $W_k$ constant and now also hold $q_k$ constant to reflect the rest
of the recovery period.  In other words, the sender is sending packets but
not increasing its window, hence the queue length should remain fixed:
\begin{equation}\label{qk4_equa}
q_{k+1} = q_k.
\end{equation}

\begin{figure}
 \begin{center}
 \scalebox{.5}{\includegraphics{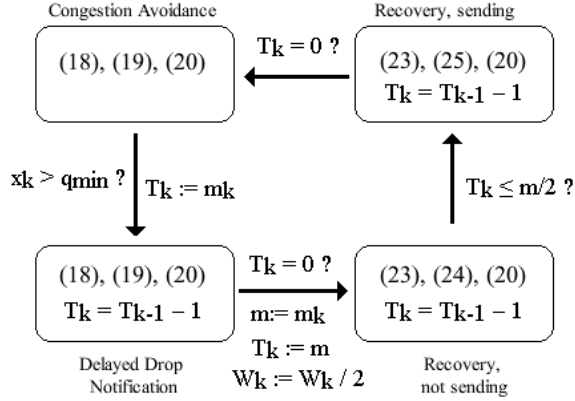}}
 \end{center}
 \caption{Hybrid systems view of the discrete-time model.}
 \label{hsD_fig}
\end{figure}

As with the continuous-time model, we can present the discrete-time model in
hybrid systems form as pictured in Fig.~\ref{hsD_fig}.  Each rounded rectangle
contains the numbers of the equations to be used, 
and a timer variable, $T_k$, is utilized.  The model begins
in the congestion avoidance phase, obeying equations~(\ref{Wk_equa})
,~(\ref{qk_equa}), and~(\ref{xk_equa}).
Once $x$ exceeds $q_{min}$, the model assumes a drop occurs, and it 
transitions to the delayed drop notification phase.  In this phase, it 
continues to obey the original set of equations,
but now it also utilizes the count-down timer, $T_k$, which expires one 
round-trip time from the time $x$ exceeds $q_{min}$.  (The round-trip
time is given by $m_k$, which was defined following
equation~(\ref{Rk_equa}).)  Once this 
timer expires, the model cuts $W_k$ in half, initializes a new count-down
timer at the value of the present round-trip time (which we label $m$) and 
transitions into the 
recovery phase in which no packets are being sent.  
In this phase of the recovery, replace equations~(\ref{Wk_equa}) 
and~(\ref{qk_equa}) used in the earlier phases, with~(\ref{Wk3_equa}) 
and~(\ref{qk3_equa}), respectively.  As mentioned above, this phase should
last $W/2$ time steps.  It is not difficult to show that this $W/2$ is
equal to $.5m$
though we leave this result out for the sake of brevity.  When the timer
has been reduced by $.5m$, the model moves into the final state -- recovery
with transmission.  Here it uses equations~(\ref{Wk3_equa}),~(\ref{qk4_equa}) 
and~(\ref{xk_equa}).  Finally, when the count-down timer expires, the model 
returns to the congestion avoidance phase.

\subsubsection{MODELING MULTIPLE SENDERS}
In order to account for multiple senders, we must modify the discrete-time 
model.  Consider the case of two senders, with window sizes of $W^{(1)}$ 
and $W^{(2)}$.  Typically the senders will alternate sending windows of 
packets.  
As a result, acknowledgments will arrive at a given sender in bunches.  
Thus, the $1/W_k$ increase per time step $\delta$ indicated in~(\ref{Wk_equa}), 
will not actually occur each time step for each sender.  However, each 
connection's window will still increase by $1$ per round-trip time.  One way 
to reflect this in the model is to choose a time step of $2\delta$, use the 
original $W$ equation~(\ref{Wk_equa}) for both senders, and change the 
queue equation to
\begin{equation}\label{qk2_equa}
q_{k+1}=q_k+\frac{1}{W^{(1)}_k} + \frac{1}{W^{(2)}_k}.
\end{equation}
For $n$ senders, use a time step of $n \delta$ and include the term  
$1/W^{(j)}_k$ in the queue equation for each sender $j=1,2,...,n$.

\subsection{RESULTS}\label{Results}
In this section we compare results obtained from applying our 
models to the network set-up mentioned above with those 
obtained using the ns simulator on an equivalent network.   We used the 
settings listed in Table~\ref{paramTable}.

%\begin{tabular}{lll}
%\textbf{Variable Name} & \textbf{Description} & \textbf{Value} \\
%\hline
%$a$ & Fixed propagation delay & .01 s \\
%$c$ & Bottleneck link capacity & 1.5 Mbps \\
%$q_{min}$ & RED parameter & 50 \\
%$q_{max}$ & RED parameter & 100 \\
%$p_{max}$ & RED parameter & .1 \\
%$w$ & RED parameter & .003 \\
%---- & Packet Size & 1000 bytes \\
%\end{tabular}

\begin{table}
\renewcommand{\arraystretch}{1.0}
\caption{Parameter Settings}
\label{paramTable}
\centering
\begin{tabular}{|c|c|c|}
\hline
\bfseries Variable & \bfseries Description & \bfseries Value\\
\hline
$a$ & Fixed propagation delay & .01 s\\
\hline
$c$ & Bottleneck link capacity & 1.5 Mbps \\
\hline
$q_{min}$ & RED parameter & 50 \\
\hline
$q_{max}$ & RED parameter & 100 \\
\hline
$p_{max}$ & RED parameter & .1 \\
\hline
$w$ & RED parameter & .003 \\
\hline
---- & Packet size & 1000 bytes \\
\hline
\end{tabular}
\end{table}

\begin{figure}
 \begin{center}
 \scalebox{.6}{\includegraphics{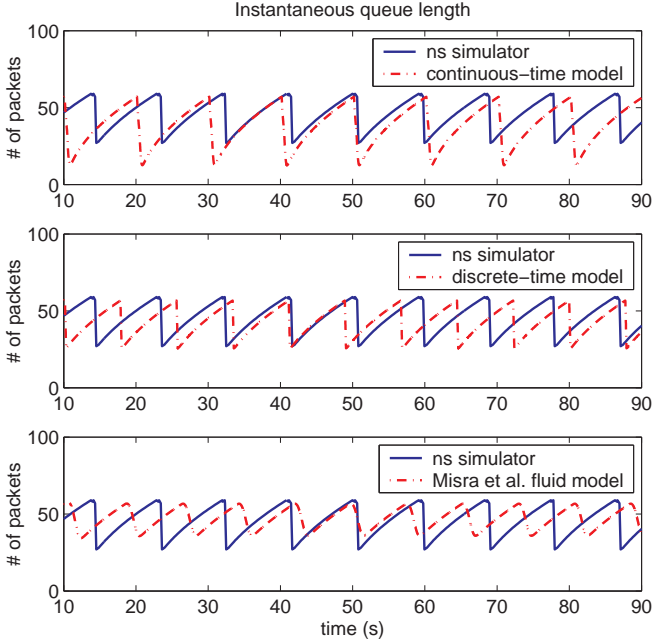}}
 \end{center}
 \caption{Comparison for one sender}
 \label{res_1}
\end{figure}

We implemented the models in MATLAB.  Since we do not model the slow-start
behavior of TCP, we begin the comparison 
after an initial transient has ended.  In Fig.~\ref{res_1}, the top plot 
shows a comparison of the queue length in the continuous-time model with the 
results from the ns network simulator.  The middle plot shows the results of
the discrete-time model compared to the simulator.
The bottom plot compares results from 
the fluid model of Misra et al.~\cite{MGT00} with the simulator~\cite{NS}. 

Despite their lack of a stochastic component as well as several of the 
details of the TCP implementation, our models show good qualitative 
agreement with the simulator.  We believe that the models have captured 
the essential behavior of this network under these flow conditions.  The 
use of an impulse helps the models account for the sharp declines in the 
queue length caused by drop events.  Lacking this feature, the fluid model 
of Misra et al. does not perform as well in this case of one sender.

One of the continuous-time model's inaccuracies is that the queue length
drops down too far.  
Lacking the no-send phase of the recovery that was included in the 
discrete-time model, the continuous-time model's queue does not drop as much 
during the recovery period.  This in turn causes $x$ to stay artificially high,
high enough so that after the recovery is complete, it is still above 
$q_{min}$.  As a result, a subsequent drop occurs, followed by an additional
window cut and recovery period, all resulting in a further drop in the queue
length.  The consecutive recoveries' negative contribution to queue length
outweigh the positive contribution on the queue length of the lack of a 
no-send phase.  The discrete-time model does not have this problem due to 
the use of the no-send recovery phase.  

Another discrepancy found between our models and the simulator is in the 
period length of the queue oscillations.  The discrete-time model's period 
is too short because of the assumption of an instantaneous drop.  At these
settings, it takes, on average, around one second from the time RED turns on
until a packet is dropped.  In Sec.~\ref{extensions} we introduce a method to estimate
this delay and incorporate it in the models.  The continuous model's period 
is too long because of the dual recovery periods which outweigh the effect 
just mentioned.

\begin{figure}
 \centering
 \scalebox{.6}{\includegraphics{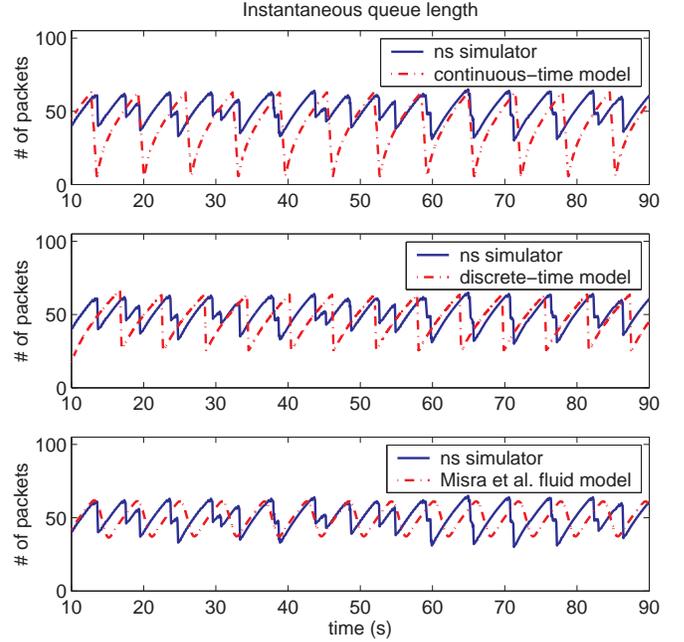}}
 \caption{Comparison for two senders}
 \label{res_2}
\end{figure}

Fig.~\ref{res_2} shows a similar comparison except now we consider 
two senders on the network with identical fixed delays. 
The network has become harder to model as evidenced by the increased 
randomness in queue variations.   Around the region $60\le t\le80$,
the simulator's behavior is analogous to its typical behavior in the 
one-sender case.  Hence, the statements made in the previous paragraph
apply there.  What is occurring in that region is one drop for 
each sender
for each drop event.  For much of the rest of the time interval shown, the
variation is due to different drop occurrences such as two drops for one sender,
only one drop, three drops and others.   This suggests a modeling approach
more directed towards statistical agreement than short-term qualitative 
agreement for cases of anything but a small number of senders.  

For larger numbers of senders the model of Misra et al$.$ shows better 
agreement with the simulator than our models do.  Thus our models are 
advantageous mainly in cases when the bottleneck link congestion is 
due to a small number of senders. 
We propose extensions to our models in the following section to improve their 
performance.

% Removed the 3,5 sender figure from here.

\section{EXTENSIONS}\label{extensions}
Model simplicity is important if we are interested in being able to concisely 
describe a system, and for amenability to mathematical analysis.  On the other
 hand, for the purpose of traffic state estimation, the key attributes we desire in 
a model are accuracy and computational efficiency.  With this in mind, we 
introduce
extensions to the models to improve their accuracy while still keeping the models 
simpler than a full-fledged packet-level simulator.   The primary modification
we consider is to implement a more realistic packet dropping mechanism within 
the models, one that more closely resembles RED.  

\subsection{ESTIMATING THE INTERDROP TIME}
As was mentioned in the previous section, one of the discrepancies between the
 models and simulator is that the models assume a drop occurs as soon as RED turns
 on.  In reality, there tends to be a delay of about one second under the 
settings we are using.  Here we describe a method to estimate this delay that 
we can incorporate into the models while still keeping them deterministic. 

In practice, RED has an added layer of complexity. 
Namely, it has been designed so that effectively, the drop time is chosen
from a uniform distribution on a time interval of length $1/p$,
given a constant RED drop probability $p$.
As implemented in the ns simulator~\cite{NS}, an additional \emph{wait} 
parameter is used, which causes this time interval
to lie between $1/p$ and $2/p$ from the time of the last drop.
(This spacing also applies to the time between RED turning on
 and the first drop.)  As a result, the expected time between drops 
is $3/(2p)$.

More precisely, when $x$ exceeds $q_{min}$ and RED turns on, 
it begins counting packets in order to determine when a 
packet drop should occur.  
The actual drop probability used by RED, which we denote $p_{drop}$, is given by:
\begin{equation}\label{drop}
p_{drop}(p,k)=
\left\{
   \begin{array}{ll}
        0 & \mbox{when $k < 1/p$}, \\
        p / (2-k p) & \mbox{when $1/p\le k < 2/p$}, \\
        1 & \mbox{when $k \ge 2/p$},
   \end{array}
 \right.
\end{equation}
where $k$ is the number of packets that have arrived since RED turned on, or 
since the previous packet drop.  Let $K$ be the random variable representing
the number of packets that arrive between drops, or between RED turning on 
and the first drop.  If $x$ is constant, then $p$ is constant, in which case
it can be shown that $K$ is uniformly distributed between $1/p$ and $2/p$.
This means $E[K]=3/2p$.  

Of course $x$ is generally not constant, and in our model we make the rough
approximation that $E[K]$ is the solution to the equation $k_{drop} = 
3/[2p(x_{k_{drop}})]$, where $x_k$ is the value of $x$ when the packet counter
reaches $k$.  We further approximate $x$ as a linear function of $k$ between
packet drops.  Rewriting~(\ref{x1_equa}) as 
\begin{equation}\label{x2}
x_k=x_{k-1}+w(q_k-x_{k-1}),
\end{equation}
we replace $q_k-x_{k-1}$ by $q_0-x_0$:
\begin{equation}\label{x2}
x_k=x_0+kw(q_0-x_0)
\end{equation}
Assuming that $x_k$ remains between $q_{min}$ and $q_{max}$,
\begin{equation}\label{pa1}
p(k) = a_1k+a_2
\end{equation}
where $a_1$ and $a_2$ are constants determined by equations~(\ref{RED_equa}) 
and~(\ref{x2}).

Since we want to find $k_{drop}$ such that $k_{drop}=3/[2p(x_{k_{drop}})]$, we substitute~(\ref{pa1})
into this equation.  Solving the resulting quadratic
produces the following solution:
\begin{equation}\label{soln}
k_{drop}=\frac{-a_2 + \sqrt{a_2^2+ 6 a_1}}{2 a_1}
\end{equation}
If $a_2^2 + 6 a_1 < 0$, then since $a_1 < 0$, $p(k)$ is a decreasing function of
$k$, and we assume that it becomes zero before another drop occurs.

\begin{figure}
 \centering
 \scalebox{.4}{\includegraphics{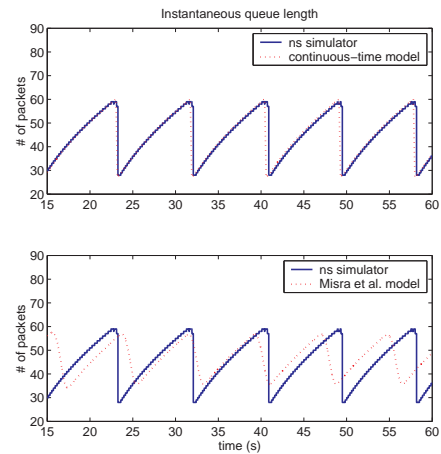}}
 \caption{Comparison of models with ns simulator for the one-sender 
scenario described in Sec.~\ref{Results}.}
 \label{tilDrop}
\end{figure}

We change the model as follows.  When RED turns on, we compute $k$, the 
expected time until the next packet drop.  A timer variable counts down this
 amount and when it expires, we consider the drop to have occurred and follow
 the same procedure as in the previous models.  This scheme can be implemented 
for both the
 continuous-time and discrete-time models.  In Fig.~\ref{tilDrop}, we present
 results of its application to the continuous-time model.  The figure shows
that using this extension, the model is able to very closely reproduce the
behavior of the simulator.  However, the accuracy of this approach diminishes
as the number of senders increases.  For even greater accuracy, we fully 
implement RED as it is executed in the ns simulator.  This is described in
the following section. 

%\begin{figure}
% \centering
% \scalebox{.4}{\includegraphics{DisVsSim5.eps}}
% \caption{Comparison for one sender between the ns simulator and the 
%discrete-time model using the interdrop time estimation.
%}
% \label{tilDrop}
%\end{figure}

\subsection{FULL RED IMPLEMENTATION}
In order to fully model RED, we begin by following the procedure laid out 
in the previous 
section up to equation~(\ref{drop}).  At that point, a uniform random number
between $0$ and $1$ is generated and if this number is less than $p_{drop}$, a 
drop occurs.  In the case of multiple senders, the probability of a given 
sender experiencing the drop is proportional to that sender's share 
$(\frac{W_i}{R_i})/(\sum_{j=1}^N \frac{W_j}{R_j})$
of the 
overall flow.  This introduces a greater degree of complexity and a stochastic
component to the model.  Since our ultimate goal is network traffic state
estimation, our main priorities are model accuracy and computational 
efficiency.

\begin{figure}
 \centering
 \scalebox{.4}{\includegraphics{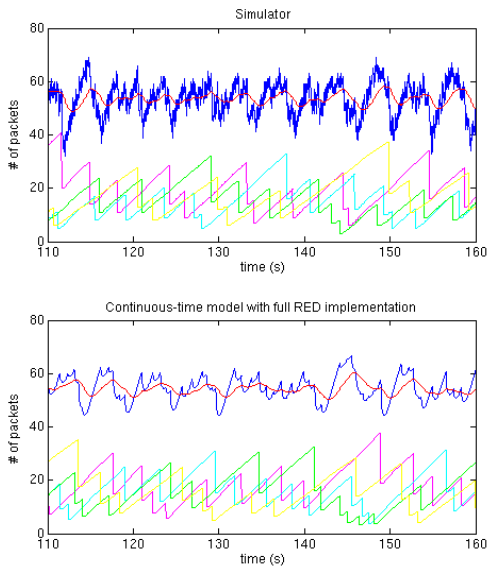}}
 \caption{Comparison for four senders between the ns simulator and 
a realization of the continuous-time model using the full RED implementation.  
In both plots,
the upper two curves represent the instantaneous and exponentially
averaged queues.  (The exponentially averaged queue is the curve showing less
variation.)  The lower four curves in both plots represent the window sizes
of the four senders.}
\label{fullRed1}
\end{figure}

\begin{figure}
 \centering
 \scalebox{.4}{\includegraphics{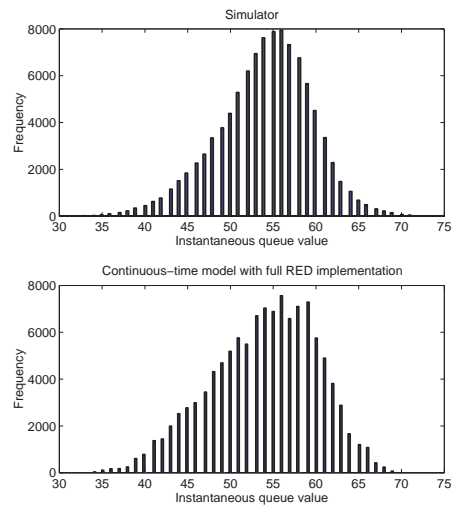}}
 \caption{Comparison of instantaneous queue histograms for the four-sender network.  The 
model is using the full RED implementation.  Model queue values, which may be fractional,
have been rounded for the purposes of comparison.}
 \label{fullRed2}
\end{figure}

We applied this model to the network described earlier, but now with four
senders at varying fixed propagation delays from the receiver.  
Fig.~\ref{fullRed1} shows that there is good qualitative agreement between
the model and the simulator.  The additional jitter in the simulator's
queue is due to packet level activity.  The model shows good statistical
agreement with the simulator as well.  Fig.~\ref{fullRed2} shows a 
comparison of queue histograms.  Note that the model
correctly captures the range in queue values, which can be used to calculate
the range of round-trip times.

In order to evaluate the model's responsiveness to network changes, we 
tested it on a network with two classes of flows, one of which turns off for part of the
run.  The classes both consist of five bulk transfer senders, but the fixed propagation delay
is different for each class: $20$ ms for class $1$ senders, $35$ ms for class $2$ senders.
The class $2$ senders turn off from $75$ to $125$ seconds after the start of the run.  In
practice, this effect could be caused by the sender-side application not having 
data to send for that stretch of time.  Another possibility could be that the class $2$ 
senders are temporarily rerouted.  

\begin{figure}
 \centering
 \scalebox{.6}{\includegraphics{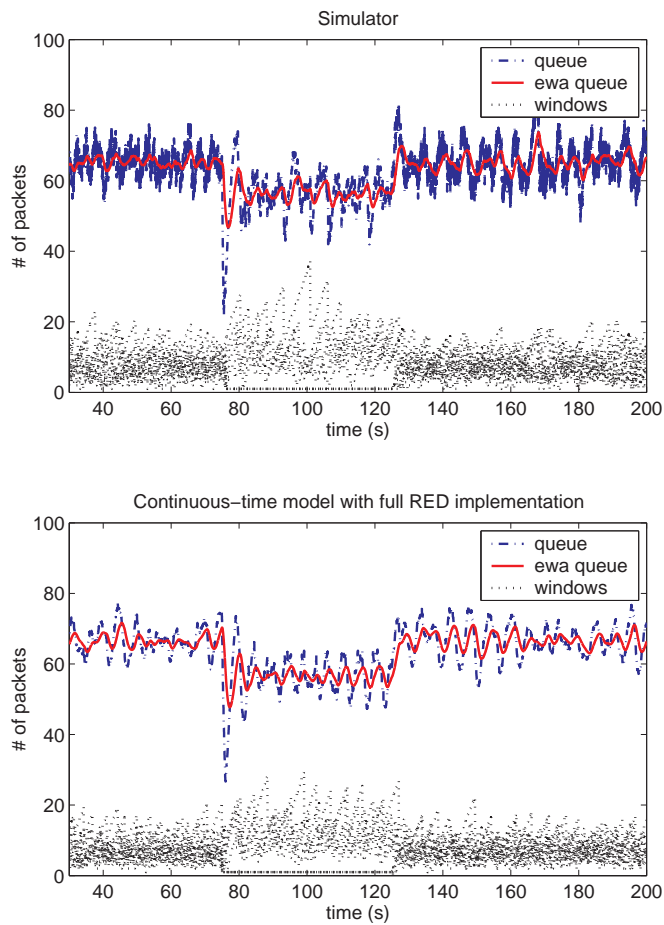}}
 \caption{Comparison for two classes with five senders per class.  Class 2 
senders are off from 75 to 125 seconds.}
 \label{10f2c}
\end{figure}

The results, shown in Fig.~\ref{10f2c}, indicate that the model does a good job of capturing the
changes in the network state caused by the senders turning off.  This includes the 
sudden drop in the queue just after the senders turn off.  
The model also reproduces the overall 
decreased level of the queue and increased level of the class 1 senders' windows during
the class 2 off period.  Since both simulator and model have stochastic components, 
each plot
represents a realization rather than the expected performance.  Nevertheless, based on our 
observations of many realizations for both model and simulator, we expect the average
 performance to show good correspondence as well.  For the time being, we give
 statistics from 
sample realizations in Table~\ref{10f2cT}, indicating the close statistical match between 
model and simulator.  Note that the model is able to estimate the variance of
the queue to within $10\%$.  This implies that by using the model in conjunction 
with the round-trip time equation~(\ref{R_equa}), 
a reasonable estimate of the variance of the 
round-trip time can be obtained.  

\begin{table}[ht]
\renewcommand{\arraystretch}{1.0}
\caption{Statistical Comparison}
\label{10f2cT}
\centering
\begin{tabular}{|c|c|c|c|}
\hline
\bfseries Variable & \bfseries Quantity & \bfseries Simulator & \bfseries Model\\
\hline
\multicolumn{4}{|c|}{All Flows On}\\
\hline
$q$ & mean & 65.5 & 66.7 \\
\hline
$q$ & st.dev. & 4.8 & 4.4 \\
\hline
\multicolumn{4}{|c|}{Half of the Flows Off}\\
\hline
$q$ & mean & 56.0 & 56.0 \\
\hline
$q$ & st.dev. & 7.0 & 6.5 \\
\hline
\end{tabular}
\end{table}

\section{TOWARDS NETWORK STATE ESTIMATION}\label{Ongoing}
Ultimately, we plan to use these models in a network traffic state estimation scheme.
The state estimation problem can be posed as follows:  Consider a network of $N$
 senders and one bottleneck router, whose state can be characterized by the sender 
window sizes and the router queue and exponentially averaged queue lengths.  
Given only a series of observations of some subset of the full network state, such
as the window sizes and round-trip times of a small number of senders, what is our 
best estimate of the state of the system at the current time?  That is, we must 
estimate all of the sender window sizes, using only the past history of some 
observed portion of the state of the network.

The estimation scheme we have in mind can be described as a particle filter or, 
more generally, a sequential Monte Carlo method~\cite{GSS93,DdFG01}.    
The basic idea
 is related to that of a Kalman filter.  But unlike the Kalman filter, the particle
 filter makes no assumptions about the linearity of the model, or about
probability distributions being Gaussian.  Thus it is a better fit for our model and system.  

The particle filter starts off with a random ensemble of states.  It uses the 
observation to filter and re-weight the ensemble members based on how 
consistent they are with the observation.  Then the ensemble members are advanced 
in time using the model, and the filter and re-weighting process repeats.  We 
are currently testing this procedure in a variety of network scenarios.

\section{CONCLUSION}

We have developed two deterministic models 
with a discrete impulse that successfully capture the network behavior of
TCP/RED in simple cases.  Extending the models by adding a stochastic 
dropping mechanism consistent with RED, the models shows good correspondence
in more complicated network situations, including those with larger numbers
 of senders and flows turning on and off.  

Aside from network traffic state estimation described in Sec.~\ref{Ongoing}, 
%other works in progress include modeling networks with multiple bottlenecks
%and cross-traffic, consideration of the physical layer, and comparisons of the
%model to real networks.
other works in progress include estimation of some dynamical 
properties such as period length, maximum sender window size and drop rate. 
% We are also considering the case of a compound bottleneck in which one 
%bottleneck link leads into a second bottleneck link of an even smaller 
%capacity.  
Additionally, we are applying our models to networks subject to
cross-traffic, including one that
is prone to irregular behavior due to the presence of a source that sends
 large bursts of packets at regular intervals.
\footnote{This work is part of the first author's Ph.D. dissertation.}


\begin{thebibliography}{9}
  \bibitem{MAF04}
     A. Medina, M. Allman and S. Floyd, ``Measuring the Evolution of Transport
     Protocols in the Internet'', preprint, available from 
     \texttt{http://www.icir.org/tbit.}
  \bibitem{FKM03}
     M. Fomenkov, K. Keys, D. Moore and k claffy, ``Longitudinal Study of 
     Internet Traffic in 1998-2003'', technical report, Cooperative Association
     for Internet Data Analysis (CAIDA), 2003.
  \bibitem{FJ93}
     S. Floyd and V. Jacobson, ``Random Early Detection Gateways for Congestion
     Avoidance,'' 
     \emph{IEEE Trans. on Networking}, Vol.1, no. 7, pp. 397-413,
     1993.
  \bibitem{MGT00}
     V. Misra, W. Gong, D. Towsley. ``Fluid-based Analysis of a Network of AQM 
     Routers Supporting TCP Flows with an Application to RED,''  
     \emph{Proc. of SIGCOMM 2000}.
  \bibitem{FB00}
     V. Firoiu and M. Borden, ``Study of Active Queue Management for 
     Congestion Control,'' \emph{Proc. of Infocom 2000}.
  \bibitem{RA01}
     P. Ranjan, E. Abed and R. La, ``Nonlinear Instabilities in TCP-RED,''   
     \emph{Proc. of Infocom 2002}.
  \bibitem{LPW02}
     S.H. Low, F. Paganini, J. Wang, S. Adlakha and J. Doyle, ``Dynamics of 
	TCP/RED and a Scalable Control,'' \emph{Proc. of Infocom 2002}.


%  \bibitem{MBDL97}
%      M. May, J. Bolot, C. Diot, B. Lyles , ``Reasons Not to Deploy RED,''
%	 Proc.
%     IWQoS'97, 1997.
%  \bibitem{LM97}
%      D. Lin and R. Morris, ``Dynamics of Random Early Detection,'' Proc. of
%     SIGCOMM, 1997.
  \bibitem{MSMO97}
      M. Mathis, J. Semke, J. Mahdavi, and T. Ott, ``The Macroscopic Behavior
     of the TCP Congestion Avoidance Algorithm,'' \emph{Computer Communications
     Review}, Vol. 27, no. 3, 1997.
  \bibitem{PFTK98}
	J. Padhye, V. Firoiu, D. Towsley, and J. Kurose, ``Modeling TCP 
	Throughput:  A Simple Model and its Empirical Validation,''
	\emph{Proc. of  SIGCOMM 1998}.
  \bibitem{HBOL01}
	J. Hespanha, S. Bohacek, K. Obraczka and J. Lee, 
	``Hybrid Modeling of TCP Congestion Control,''
	\emph{Lecture Notes in Computer Science},
	 no. 2034, pp. 291-304, 2001.
 \bibitem{BHL03}
	S. Bohacek, J. Hespanha, J. Lee and K. Obraczka,
	``A Hybrid Systems Modeling Framework for Fast and Accurate Simulation
	 of Data Communication Networks,'' \emph{Proc. of SIGMETRICS 2003}.
 \bibitem{LLM03}
	Y. Liu, F. Lo Presti, V. Misra, D. Towsley and Y. Gu, ``Fluid Models
	and Solutions for Large-Scale IP Networks,'' 
	\emph{Proc. of SIGMETRICS 2003}.
%  \bibitem{PF01}
%	J. Padhye and S. Floyd, ``On Inferring TCP Behavior,'' 
%	\emph{Proc. of SIGCOMM 2001}.
%  \bibitem{NN02}
%        S. Northcutt, J. Novak,
%        \emph{Network Intrusion Detection}
%        Que, Indianapolis, 2002.
%  \bibitem{Ba01}
%        C. Barakat,  
%        ``TCP/IP Modeling and  Validation'',
%        \emph{IEEE Network}, (May-June 2001), 38-47.
%  \bibitem{AK99}
%        M. Altman, J. Kones,  
%        ``On the Effective Evaluation of TCP,''
%	\emph{Computer Communications Review}, Oct 1999.
   \bibitem{NS}
        \texttt{http://www.isi.edu/nsnam/ns/}
   \bibitem{GSS93}
        N. Gordon, D. Salmond and A. Smith, ``Novel approach to nonlinear/non-Gaussian 
Bayesian state estimation,''
        \emph{IEE Proceedings-F}, Vol. 140, no. 2, pp. 107-113, 1993.
\bibitem{DdFG01}
        A. Doucet, N. de Freitas, N. Gordon, eds., 
        \emph{Sequential Monte Carlo Methods in Practice},
        Springer, New York, 2001. 
\end{thebibliography}
\end{document}